\documentclass{elsarticle}

\usepackage{amstext}
\usepackage{amssymb}

\newcounter{bla}
\newenvironment{refnummer}{%
\list{[\arabic{bla}]}%
{\usecounter{bla}%
 \setlength{\itemindent}{0pt}%
 \setlength{\topsep}{0pt}%
 \setlength{\itemsep}{0pt}%
 \setlength{\labelsep}{2pt}%
 \setlength{\listparindent}{0pt}%
 \settowidth{\labelwidth}{[9]}%
 \setlength{\leftmargin}{\labelwidth}%
 \addtolength{\leftmargin}{\labelsep}%
 \setlength{\rightmargin}{0pt}}}
 {\endlist}

\newcommand{\pd }{Pad\'{e} }

\begin{document}
\begin{frontmatter}

\title{Extracting resonance poles from numerical scattering data:
type-II \pd reconstruction}

\author{D. Sokolovski$^{a,c,d}$, E. Akhmatskaya$^{b,c}$ and S.K.Sen$^{d}$}



\address[a]{Department of Physical Chemistry, University of the Basque Country, Leioa, 48940, Spain}

\address[a]{Basque Center for Applied Mathematics (BCAM),\\ Building 500, Bizkaia Technology Park E-48160, Derio, Spain}

\address[a]{IKERBASQUE, Basque Foundation for Science, E-48011 Bilbao, Spain}

\address[d]{School of Mathematics and Physics\\
            Queen's University Belfast,\\
	    Belfast BT7 1NN, United Kingdom}

\begin{abstract}
 We present a  FORTRAN 77 code for evaluation of resonance pole positions and residues
of a numerical scattering matrix element in the complex energy (CE) as well as in the complex
angular momentum (CAM) planes. Analytical continuation of the $S$-matrix element is 
performed by constructing a type-II  \pd approximant from given physical values [Bessis {\it et al} (1994); Vrinceanu {\it et al} (2000); Sokolovski and Msezane (2004)] .
The algorithm involves iterative 'preconditioning' of the numerical data 
by extracting its rapidly oscillating potential phase component.
The code has the capability of adding non-analytical noise to the numerical data in order to   
select 'true' physical poles, investigate their stability and evaluate the accuracy of the reconstruction. It has an option of employing multiple-precision (MPFUN) package [Bailey (1993)] developed by D. H. Bailey wherever double precision calculations fail due to a large number of input partial wave (energies) involved. The package has been successfully tested on several models, as well as the F + H$_2$
  $\rightarrow$ HF + H, 
   F + HD$\rightarrow$ HF + D,
  Cl + HCl$\rightarrow$ ClH + Cl and 
  H + D$_2$$\rightarrow$ HD + D reactions.
  Some detailed examples are given in the text.

\begin{flushleft}

\end{flushleft}


\end{abstract}

\end{frontmatter}


{\bf PROGRAM SUMMARY}

\begin{small}
\noindent
{\em Manuscript Title:} Extracting resonance poles from numerical scattering data:
type-II \pd reconstruction.               \\
{\em Authors:} D. Sokolovski, E.Akhmatskaya and S.K.Sen                 \\
{\em Program Title:} \texttt{PADE\_II}                      \\
{\em Journal Reference:}                                      \\
{\em Catalogue identifier:}                                   \\
{\em Licensing provisions:}   Free software license                                \\
{\em Programming language:} FORTRAN 77                        \\
{\em Computer:} Any computer equipped with a FORTRAN 90
compiler\\
{\em Operating system:} UNIX, LINUX                           \\
{\em RAM:} 256 Mb                                       \\
{\em Has the code been vectorised or parallelised:} no                                      \\
{\em Number of processors used:} one                          \\
{\em Supplementary material:} MPFUN package, validation suite, script files, input files,
readme file, Installation and User Guide
{\em Classification:} Molecular Collisions                    \\
{\em External routines/libraries:} NAG Program Library \\ 
{\em CPC Program Library subprograms used:}   N/A          \\
{\em Nature of problem:}\\
 The package extracts the positions and residues of resonance poles 
 from numerical scattering data supplied by the user.
 The data can then be used for quantitative analysis of interference patterns observed in elastic, inelastic and reactive integral and differential cross sections.
 \\
{\em Solution method:}\\
 The $S$-matrix element is analytically continued in the complex plane 
 of either energy or angular momentum with the help of \pd 
 approximation of type II. Resonance (complex energy or Regge) poles are identified and their 
 residues evaluated. 
   \\
{\em Restrictions:}\\
None.\\
  {\em Unusual features:}\\
  Use of multiple precision $MPFUN$ package.   \\
{\em Additional comments:} none\\
{\em Running time:}\\
from several seconds to several minutes depending on the precision level chosen and the number of iterations performed.\\
{\em References:}
\begin{refnummer}
\item D. Bessis, A. Haffad, and A. Z. Msezane,
Phys. Rev. A., {\bf 49} (1994) 3366.
\item D. Vrinceanu, A. Z. Msezane, D. Bessis, J. N. L. Connor and 
 D. Sokolovski, Chem. Phys. Lett., {\bf 324}, (2000) 311.
 \item D. Sokolovski, A. Z. Msezane,  
Phys. Rev. A., {\bf 70} (2004) 032710.
\item D. H. Bailey, Algorithm 719, "Multiprecision translation and execution 
of Fortran programs". ACM Transactions on Mathematical Software, 
19(3), (1993) 288.
\end{refnummer}

\end{small}

\newpage


\hspace{1pc}
{\bf LONG WRITE-UP}

\section{Introduction}

In the last fifteen years the progress in crossed beams experimental techniques has been matched by the  development  of state-of-the-art computer codes capable of modelling atom-diatom 
elastic, inelastic and reactive integral (ICS) and differential (DCS) cross-sections  to a very high accuracy \cite{Skod, CLARY98,SKOU99,CASA99}. Both quantities, often structured, carry a large amount of useful information about the details of the scattering mechanism which makes further detailed analysis of the scattering data highly desirable.
There are, in general, two distinct types of collisions: in a {\it direct} collision the partners part soon after the first encounter, while in a  {\it resonance} collision they form an intermediate complex 
(quasi-molecule) which then breaks up into products (reactive case) or back into reactants (elastic or inelastic case). Often a direct and several resonance processes are possible and,
in accordance with the rules of quantum mechanics, interference between them gives a complicated shape to an ICS or (and) a DCS.
 A collision is described by a complex valued unitary scattering $(S-)$ matrix, whose elements
 are the probability amplitudes for the transitions between all initial and final states of the collision partners. The dimensions of an $S$-matrix are equal to the number of open channels (e.g., $1\times 1$ in the simplest case of potential scattering) and in the following we will 
 consider one scattering element at a time, as the development of an efficient algorithm for analytical
 continuation of the entire matrix remains an open problem.
Mathematically, a resonance is associated with a pole of an $S$-matrix element. There are two types of such poles, closely related to each other.

 {\it Complex energy poles.} Consider a state-to-state scattering matrix element $S_{\nu' \leftarrow \nu}(E,J)$, where $E$ is the energy, $J$ is the total angular momentum and $\nu$ and $\nu'$ denote the set of quantum numbers describing the state of partners before and after the collision (e.g., 
 helicity and the vibrational and rotational quantum numbers for an atom-diatom reaction).
 In the presence of several long-lived intermediate states (a sharp resonances) $S_{\nu' \leftarrow \nu}(E,J)$ has poles at $E=E_n(J)$, $n=1,2,...$, close to the real axis in the fourth quadrant of the complex $E$-plane, i.e.,
\begin{equation}\label{1}
S_{\nu' \leftarrow \nu}(E_n,J)=\infty, \quad Re E_n>0, \quad Im E_n<0
\end{equation}%

 {\it Complex angular momentum (Regge) poles.} Conversely, one can fix the value of energy 
 $E$ and consider $S_{\nu' \leftarrow \nu}(E,J)$ as a function of the (continuos) variable $J$.
 If so, a sharp resonances would manifest themselves as poles at $J=J_n(E)$, $n=1,2,...$, close to the real axis
 in the first quadrant of the complex angular momentum (CAM) plane,
\begin{equation}\label{2}
S_{\nu' \leftarrow \nu}(E,J_n)=\infty, \quad Re J_n>0, \quad Im J_n>0.
\end{equation}%
 
Far from being just a mathematical abstraction, poles of the $S$-matrix element provide a practical tool for a quantitative analysis of the direct and resonance contributions to both ICS and DCS.
For this purpose the Regge poles are more useful then their complex energy counterparts
(for applications of Regge poles to molecular collisions see  \cite{CON0}-\cite{CON10}).
For example, an inelastic or reactive DCS is given by the square of the probability amplitude for scattering at an the angle $\theta$, $\sigma_{\nu^{\prime} \gets \nu}(E,\theta) = |f_{\nu^{\prime} \gets \nu}(E,\theta)|^2$.
For zero initial and final helicities, the scattering amplitude can be written as a partial wave sum (PWS) over all physical (i.e., non-negative integer) values of $J$ in the form 
\begin{equation}\label{3}
 f_{\nu^{\prime} \gets \nu}(E,\theta) = 
 (ik_{\nu})^{-1}\sum_{J=0}^{\infty}(J+1/2)P_{J} (cos(\theta))
 S_{\nu^{\prime} \gets \nu}(E,J)
\end{equation}%
where $P_{J} (cos(\theta))$ is the Legendre polynomial and $k_{\nu}$ is the initial wave vector.
Similarly, the  ICS at an energy $E$, $\sigma_{\nu^{\prime}}(E)$ is given by another sum over angular momenta,
\begin{equation}\label{4}
 \sigma_{\nu^{\prime} \gets \nu}(E) = 
 (\pi/k_{\nu}^{2})\sum_{J=0}^{\infty}(2J+1)
 |S_{\nu^{\prime} \gets \nu}(E,J)|^2.
\end{equation}%
By using the Poisson sum formula \cite{Brink} one can convert either sum into a sum containing integrals over 
the (continuous)  variable $J$  and transform the contour of integration so as to pick up the resonance contributions of the Regge poles $J_n$ in the first quadrant of the  CAM plane.
The rest of the integral is usually structureless and can be attributed to the direct scattering
mechanism \cite{FOOT}. Such a technique for analysing differential cross sections has been developed 
in \cite{G1}-\cite{IH2}.
A method for analysing integral cross sections was introduced in \cite{ICS1} and extended to reactive collisions in \cite{ICS2,ICS3}. Both approaches require knowing Regge pole positions $J_n$ as well as the corresponding residues $\rho_n^{\nu^{\prime} \gets \nu}$,
\begin{equation}\label{5}
 \rho_n^{\nu^{\prime} \gets \nu}(E) = lim_{J \rightarrow J_n }(J-J_n) 
 S_{\nu^{\prime} \gets \nu}(E,J).
\end{equation}%
The purpose of this paper is to propose a simple tool for evaluation of Regge pole
positions and residues from the numerical scattering data obtained in modelling realistic atomic and molecular systems. A prototype of the program described below has been successfully applied to  
the analysis of the 
  Cl + HCl$\rightarrow$ ClH + Cl reaction \cite{HCL1,HCL2},
the F + H$_2$ $\rightarrow$ HF + H  reaction \cite{FH1}-\cite{FH5},\cite{ICS2,ICS3},
 the H + D$_2$$\rightarrow$ HD + D  reaction \cite{HD1,HD2},
the I + HI$\rightarrow$ IH + I  reaction \cite{G2,IH1,IH2},
 and the   F + HD$\rightarrow$ HF + D reaction \cite{FHD1}.

\section{Type-II \pd reconstruction of  a scattering matrix element}

For a simple one-channel problem, one can obtain the Regge pole positions and residues directly by integrating the  Schroedinger equation for complex values of $J$  \cite{SE1,SE2}.
This is, however, not the case when the number of open channels and, therefore, the dimension of the $S$-matrix is large. A typical scattering code  \cite{CLARY98,SKOU99,CASA99} calculates a discreet set $S_{\nu^{\prime} \gets \nu}(E,J)$ for the physical values $J=0,1,...N$, with $N$ chosen sufficiently large for the PWS (\ref{3}) and (\ref{4}) to converge.
Similarly, one can obtain the values $S_{\nu^{\prime} \gets \nu}(E^j,J)$ 
for a fixed integer $J$ on an energy grid $E^j$, $j=1,2,...N$, where the values of $E^j$ can, unlike the those of $J$, be chosen arbitrarily.
One can, in principle, use the first set of values to obtain Regge pole positions and residues and the second to find the CE poles.

{\it Regge poles positions and residues.} 
If we are interested in the CAM poles of the $S$-matrix element at a given energy $E$, it is natural to seek its analytic continuation in the form of a ratio of two polynomials of $J$, $P(J)/Q(J)$. 
 If successful, we can expect at least some of the complex zeroes of $Q(J)$ to coincide with the true physical poles of the problem.
It is convenient to choose the orders of $P(J)$ and $Q(J)$ close to $N/2$ (see below)
and require that the ratio equals the input values of the $S$-matrix element, $S_{\nu^{\prime} \gets \nu}(E,J)$, at $J=0,1,...N$.
Such a procedure is well known in the literature  \cite{PAD1,PAD2}  and an approximant constructed in this manner is called \pd approximant of type II. A detailed description of the algorithm,
inspired by the work of Stieltjes, can be found in \cite{PAD1}. The method involves constructing a continued fraction whose coefficients $\phi_j$, $j=1,2,..N$  are determined by a recursive procedure. Knowing the set of $\phi_j$ allows one to construct the coefficients of  $P(J)$ and $Q(J)$ and, from them,
the poles and zeroes of the approximant. The order of $P(J)$ is $[N/2]$ and that of $Q(J)$ 
$[(N-1)/2]$, where $[A]$ denotes the integer part of $A$.
Although simple, the just described procedure gives no clue to how exactly the \pd approximation
would dispose of $[(N-1)/2]$ poles available to it, for example, when there is only one true resonance Regge pole. It has been shown in Ref.\cite{G1,G2,G3} that. typically, the poles as well as zeroes fall into three different categories:

\textit {(i) True poles and zeros} of the $S$-matrix element, which are 
stable with respect to number of input data points and amount of non-analytical noise present in the input data. True poles and the corresponding residues 
are the quantities of interest.

\textit {(ii) Froissart doublets.} These are zero-poles pairs placed near the real $J$ axis.
The doublets appear as the approximant tries to reproduce non-analytical noise 
 present in the input data \cite{PAD1}).

\textit {(iii) Background (border) poles and zeros.}
These form a border of a subset $\Omega$ of the CAM plane inside which the \pd approximant 
faithfully reproduces the analytic function specified by the input values and beyond which the approximant fails \cite{G1,G3}.
The size of the $\Omega$ depends on whether the approximated function has an oscillatory components (see below) and on the amount of non-analytical noise present in the input values.
Positions of the border poles and zeroes are not necessarily stable. With the increasing amount of noise, the border shrinks
and becomes less well defined, as both poles and zeroes leave it to form Froissart doublets near the real axis. 

Such a behaviour is shown in Fig.1 for a model system studied earlier in \cite{G1}.
In order to maximise the domain of validity of an approximant, it is beneficial to remove all rapidly oscillating factors. For a heavy (semiclassical) atom-diatom system, $S_{\nu^{\prime} \gets \nu}(E,J)$ typically contains a rapidly oscillating factor of the form $\exp[i\phi(J)]$ where (the constant term $c$ is unimportant, but is kept for consistency)
\begin{equation}\label{6}
\phi(J)=J^2+bJ+c,
\end{equation}%
often referred to as potential phase. Its origin was discussed in  \cite{G1}:
$d\phi(J)/dJ=2aJ+b$ represents the 'deflection function', i.e., the scattering angle
for a classical trajectory with a total angular momentum $J$. Since reaction probability 
decreases rapidly with $J$
one can replace  $d\phi(J)/dJ$ by a linear and $\phi(J)$ by a quadratic form, respectively. Note that the coefficients $a$, $b$, and $c$ are not known {\it apriori} and must 
be determined iteratively (see below). A good initial guess is $b\approx \pi$ and $a \approx - \pi/N$, which 
corresponds to a mostly repulsive collision in which trajectories with small angular momenta (impact parameters) are scattered backwards, and those with large ones, $J\approx N$, are scattered in the forward direction.
Thus a  \pd approximant takes the form
\begin{eqnarray}\label{7}
  S_{\nu' \gets \nu}^{Pade}(E,J)\equiv
 K_{N} \textrm{exp}[i(aJ^{2}+ bJ+c)]
 \times \frac{\prod_{i=1}^{[N/2]}(J-Z_i)}
 {\prod_{i=1}^{[(N-1)/2]}(J-P_{i})}, \\
 \nonumber
 S_{\nu' \gets \nu}^{Pade}(E,J)=S_{\nu' \gets \nu}(E,J), \quad J=0,1,...N,
\end{eqnarray}%
completely specified by the sets of zeroes $\{ Z_i\}$ and poles $\{P_i\}$ and the constants 
$a$,$b$, $c$ and $K_N$. The residue at the $n$-th pole in Eq.(\ref{7}), $J_n$, is explicitly given by
\begin{equation} \label{8}
  \rho^{\nu' \gets \nu}n(E)=
 K_{N} \textrm{exp}[i(aP_n^{2}+ bP_n+c)]
 \times \frac{\prod_{i=1}^{[N/2]}(P_n-Z_i)}
 {\prod_{i\ne n}^{[(N-1)/2]}(P_n-P_{i})},
\end{equation}%

{\it Selection of true poles.Non-analytical noise.} 
As discussed above, not all poles $\{P_i\}$ coincide with the true physically important poles
$J_n$ of $S_{\nu' \gets \nu}$ and one needs to choose among them. Important resonance poles are typically located above the real axis in the region containing the input values of the angular momentum. They are usually easily distinguished from both Froissart doublets and the border poles (see Fig. 3a). As true poles are expected to be stable with respect to a non-analytical noise,
an additional test can be provided by contaminating the original data with such a noise and then
selecting the poles not affected by the contamination (see Fig.3b).

{\it Complex energy poles.} Most of the above equally applies to constructing a \pd approximant 
on a grid of energy values $E^j$, $j=1,,..N$ for a given physical value of the angular momentum J. Although without a clear physical meaning, a quadratic phase similar to one in Eq.(\ref{6}) could still be extracted and one obtains
\begin{eqnarray}\label{9}
  S_{\nu' \gets \nu}^{Pade}(E,J)\equiv
 K_{N} \textrm{exp}[i(aE^{2}+ bE+c)]
 \times \frac{\prod_{i=1}^{[N/2]}(E-Z_i)}
 {\prod_{i=1}^{[(N-1)/2]}(E-P_{i})}, \\
 \nonumber
 S_{\nu' \gets \nu}^{Pade}(E^j,J)=S_{\nu' \gets \nu}(E^j,J), \quad j=0,1,...N.
\end{eqnarray}%
Again, several of the poles $\{P_i\}$ may correspond to the true CE poles of a given partial wave, while the rest would belong to the border or form Froissart doublets (Fig.5). The true poles can be selected either on physical grounds or by contaminating the data with 
additional noise as discussed above.

{\it Multiple precision.} Finally, when the number of input partial waves (energy points) $N$ becomes large  ($\sim 100$), construction of a \pd approximant may require evaluation of a polynomials of  orders  so high that the  double precision would be insufficient.  Reference \cite{PAD1} estimates
the required accuracy of $2N$ decimal points for $N$ input values. We found this condition to be too restrictive, and an example when multiple precision calculations are necessary will be given in the following  (see Fig.6).

\section{Description of the code}

The
\textit{PADE\_II.f} application is  a sequence of 27 FORTAN 77 files. 
In addition,
the Multiple Precision Floating Point Computation Package (MPFUN) 
\cite{MPFUN} developed by D. H. Bailey and two utility programs,
 the translator  ({\it transmp90.f}) and the validating utility 
({\it validate.f}) are supplied.
All the relevant documentation and procedure specifications are 
provided within the program codes. 

\section{Using MPFUN}

At the beginning of each FORTRAN 77 file to be translated to a higher precision level,
the directives (i.e., special comments) are inserted in the form
\begin{verbatim}
                CMP+ PRECISION LEVEL 60
                CMP+ OUTPUT PRECISION 60.
\end{verbatim}
This determines the number of digits after the decimal point 
to be used in a calculation.
It has been found that  double precision (DP) is sufficient 
in calculations with $\lesssim 50$ input points. 
 The variables in the main program or 
in a subprogram to be converted to 
 multiple precision (MP) by the translator program are declared 
by explicit MP type directives, such as:
\begin{verbatim}
                 CMP+ MULTIP REAL
                 CMP+ MULTIP COMPLEX. 
\end{verbatim}		 
One must also specify whether a constant in the input program 
will be treated as an MP quantity. This is done 
by appending to the constant a flag $+0$, as in the following examples:
\begin{verbatim}
                1.2345678553884665E-13+0
                pi=acos(-1.0+0)
                zi=dpcmpl(0.+0,1.+0)
\end{verbatim}
For a more detailed discussion of the multiple precision options
 we refer the reader to Ref. \cite{MPFUN}.
 We note that MP is always used in the intermediate calculations and only in some
parts of the program (for example, the NAG routines cannot be converted to MP). 
Thus the output is always given in the DP format regardless
of which level of accuracy has been used in the intermediate steps.

\section{Program structure of \textit{PADE\_II} (\textit{PADE\_II.f})}
The main program contains a cycle of $niter$ steps,
 in which the potential phase (\ref{6}) is iteratively removed
from the input data and a new set of poles and zeroes is recalculated at each step.
If one wishes to do so, the iteration cycle can be repeated $nstime$ times, each time with
a different random noise to check stability of the poles, as has been discussed above.
Parameters $niter$, $nstime$ and the magnitude of the noise are supplied by the user 
and defined in the input files as discussed in the following. Given below is the calling sequence of the subroutines. 
\begin{verbatim}
                call open_files
     666        call noise
                call precond
     777        if(imult.eq.1)then
                call dptomp
                if(iexp.eq.1) call mulexp
                call pade1_mp
                call pade2_mp
                call findzeros_polymp
                call findpoles_polymp
                else
                call pade1
                call pade2		
                call findzeros_nag
                call findzeros_poly
                call findpoles_nag
                call findpoles_poly
                endif
                call smooth
                call fit
                if (icheck.le.niter) go to 777
                call spade
                call residue
                call output
                if(nscheck.le.nstime) go to 666.
\end{verbatim}		
Here we have suppressed arguments of the subprograms described in detail in the following Section.

\section{Subroutine specifications} 

\subsection{Subroutine \texttt{OPEN\_FILES} (\texttt{open\_files.f})}

Two input files are employed in the subroutine \texttt{OPEN\_FILES}, 
assigning several parameters/variables with their values and 
the input data for which \pd reconstruction is to be performed.
\newline 
Parameters listed in the input file, 'unit 1', (the name of this input file should be specified by the user, as desrcibed in Sect. 9) are as follows:

\noindent
\textbf {nread}: number of input data points available in 'unit 1'.\\

\textbf {niter}: number of iterations required for the convergence of the calculation\\

\textbf {shift}: this shifts the input grid points and may be used to
avoid exponentiation of extremely large number when evaluating the polynomials involved. 
For Regge poles calculations the value $nread/2$ is suggested.\\

\textbf {jstart} and \textbf {jfin}: with all input points numbered by $j$ between $1$ and $N$,
determine a range $jstart \le j \le jfin$ to be used for a \pd reconstruction.
This gives the user an additional flexibility. The values $jstart=1$ and $jfin=N$ are recommended.\\

\textbf {inv}: set to '-1' and not used in the current version of the program \\

\textbf {dxl}: defines a strip $-dxl < Im J (Im E) < dxl$ such that all poles and
zeroes within the strip will be removed when calculating the potential phase
of the approximant.\\

%

Parameters listed in the input file \textit {param.pade}, 'unit 2',
are as follows:

\noindent
\textbf {ipar}:  has value either '0' or '1', depending on whether
$S$-matrix element do not require or require parity inversion,
$S_{\nu' \leftarrow \nu}(E_n,J)\rightarrow \exp(i\pi J)S_{\nu' \leftarrow \nu}(E_n,J)$.
Different codes calculating $S$-matrix elements use different conventions,
and a parity change may be necessary to bring them to a single standard. \\

\textbf {iprec}: has values of either '1' or '0' depending on whether or not the input data
is to be 'preconditioned' (see below) prior to the start of a calculation.\\

\textbf {imult}: has values of either '1' or '0'  depending on whether a calculation is
to be performed with multiple or double precision.\\

\textbf {nstime}: number of times user would like to contaminate input 
data with non-analytical noise. Addition of such noise (apart from those 
present in the numerical data) is done in the loop labelled as '666' 
(see section 6). 'nstime', initially assigned to '0', indicates 
input data is not contaminated with random noise for the first 
set of iterations.\\

\textbf {nprnt}: number of points used in plotting the resultant
\pd approximant for real values of the argument.\\

\textbf {fac}: determines the magnitude of the noise added to the initial data.\\
The input parameters are passed between subroutines via common statements,
\begin{verbatim}
         common/print/nprnt
         common/para/niter,shift,jstart,jfin,inv,dxl
         common/par/ipar,iprec,imult,nstime,nscheck
         common/nsch/fac,nread
\end{verbatim}



The output contains \texttt{n = jfin - jstart + 1} values of the argument,
$tt(i)$, and the function, $zfk(i)$, to be used for \pd reconstruction,

\begin{small}
\begin{tabular}{l l l l}
\hline
output & tt(*)  & DP, real    & angular momentum, energy \\
output & zfk(*)  & DP, complex & $S$-matrix \\
\hline
\end{tabular}
\end{small}

Here and in the following a comment $(*)$ indicates that a variable is passed to 
the main program or other subroutines through a $COMMON$ statement.  

\subsection{Subroutine \texttt{NOISE}(\texttt{noise.f})}
This adds random noise to the input data, 
\begin{equation} \label{noise}
  S(J) \rightarrow S(J) + fac \times (g_{1} + i g_{2})
\end{equation}%
where $g_{1}$ and $g_{2}$ are real random variables taking values 
between $-1/2$ and $1/2$ and the noise magnitude $fac$ is specified by the 
user in the input file \textit {param.pade}. Random numbers are generated
in \texttt{zsrnd} function (\texttt{zsrnd.f})
 using the 
\texttt{G05CAF} subroutine of NAG program library \cite{NAGNOISE} 
designed to generate pseudo-random real numbers, distributed 
between (0,1). The input and output are:

\begin{small}
\begin{tabular}{l l l l}
\hline
input  & zfk   & DP, complex & $S$-matrix element      \\
output & zfkns & DP, complex & $S$-matrix with random noise added. \\
\hline
\end{tabular}
\end{small}

\subsection{Subroutine \texttt{PRECOND}(\texttt{precond.f})}

\begin{small}
\begin{tabular}{l l l l}
\hline
input  & tt (*)   & DP, real    & input grid values \\
input & zfkns & DP, complex & $S$-matrix (with random noise) \\
output & ts   & DP, real    &shifted  input grid values \\
output & zsm  & DP, complex & preconditioned $S$-matrix \\
\hline
\end{tabular}
\end{small}

 This subroutine prepares input data
according to the values of 'ipar', 'iprec' and 'shift'.

If 'ipar=1' input data is given an additional phase,
%
${S}(J) \rightarrow S(J)\textrm{exp}(i\pi J).$%

If 'iprec=1' a quadratic phase 
$\Phi(J)= \Phi_2 J + \Phi_3 J^2$  with  $\Phi_2=\pi$ and 
$\Phi_3=\pi /(2 (n-1))$, is removed from the input values,
${S}(J) \rightarrow S(J)\textrm{exp}[-i\Phi(J)].$.

If the value of 'shift' is not zero, the grid points are shifted,
\texttt{ts(i)=tt(i)-shift.}  

If ipar=0, iprec=0, shift=0, the input data  
remains unchanged. 
	       
\subsection{Subroutine \texttt{DPTOMP}(\texttt{dtomp.f})}
 This subroutine converts DP data to 
an MP format. We will follow the convention that if 
the name of a variable ends in \texttt{'mp'} the variable 
may be converted to MP.

\begin{small}
\begin{tabular}{l l l l}
\hline
input  & ts   & DP, real    & angular momentum/energy \\
input & zsm & DP, complex & preconditioned $S$-matrix \\
output & tsmp   & MP, real    & angular momentum/energy \\
output & zsmmp  & MP, complex & preconditioned $S$-matrix \\
\hline
\end{tabular}
\end{small}

If the parameter 'imult' is set to '0' in the input file {\it param.pade}, the calculation
will be done in double precision and will not involve MP subroutines. 

\subsection{Subroutines \texttt{PADE1}(\texttt{pade1.f})/ \texttt{PADE1\_MP}(\texttt{pade1\_mp.f})}

\begin{small}
\begin{tabular}{l l l l}
\hline
input & ts/tsmp & DP/MP, real & shifted angular momentum, energy\\
input & zsm/zsmmp & DP/MP, complex & preconditioned data\\
output & zphi/zphimp & DP/MP, complex & coefficients of the continued fraction \\
\hline
\end{tabular}
\end{small}

Both subroutines evaluate,
following the method detailed in \cite{PAD1},
coefficients $\phi_{i}$ of continued fraction 
\begin{eqnarray}\label{frac}
F_{n}(t) = \phi_{1} + \frac {t-ts(1)}{\phi_{2}+\frac {t-ts(2)}{\phi_3+ 
          \frac {t-ts(3)}{\ddots
	  +\frac {.}{\phi_{n-2}+\frac {t-ts(n-2)}{\phi_{n-1}+
	  \frac{t-ts(n-1)}{\phi_n}}}}}}\\
	  \nonumber
\end{eqnarray}%
subject to the condition that at the grid points the value $F_{n}(t)$ coincides with 
the corresponding input values, $F_n[ts(i)]=zsm(i)$, $i=1,2,..n$.

%
%
%
A calculation employs 
\texttt{PADE1\_MP} for an MP calculation ( \texttt{'imult=1'})
or $ \texttt{PADE1}$ for a DP calculation  ( \texttt{'imult=0'}).

\subsection{Subroutines \texttt{PADE2}(\texttt{pade2.f})/ \texttt{PADE2\_MP}(\texttt{pade2\_mp.f})}
With  $[A]$ denoting the integer part of a number $A$,
the continued fraction $F_{n}(t)$ can be written as a ratio of two polynomials $P_n(t)/Q_n(t)$ of 
the degrees $[n/2]$ and $[(n-1)/2]$, respectively,
\begin{eqnarray}
P_{n}(t)=\sum_{j=0}^{[n/2]} p_j t^j, \quad
 \nonumber
Q_{n}(t)=\sum_{j=0}^{[(n-1)/2]} q_j t^j.
\end{eqnarray}
Subroutine $\texttt{PADE2}$$/\texttt{PADE2\_MP}$ evaluates the coefficients 
$p_j$ and $q_j$ recursively, using the method detailed in  \cite{PAD1}.

\begin{small}
\begin{tabular}{l l l l}
\hline
input & ts/tsmp & DP/MP, real & angular momentum/energy\\
input & zphi/zphimp & DP/MP, complex & rational fraction \\
output & kp(*) & integer & degree of $P_n(t)$\\
output & kq(*) & integer & degree of $Q_n(t)$\\
output & zpn/zpnmp & DP/MP, complex & coefficients of $P_n(t)$\\
output & zqn/zqnmp & DP/MP, complex & coefficients of $Q_n(t)$\\
\hline
\end{tabular}
\end{small}

A calculation employs 
\texttt{PADE2\_MP} for MP calculations ( \texttt{'imult=1'})
or $ \texttt{PADE2}$ for DP calculations ( \texttt{'imult=0'}).

\subsection{Subroutines\newline  \texttt{FINDZEROS\_POLY}(\texttt{findzeros\_poly.f}),\newline
 \texttt{FINDZEROS\_POLYMP}(\texttt{findzeros\_polymp.f}),
   \newline \quad \texttt{FINDPOLES\_POLY}(\texttt{findpoles\_poly.f}), 
   \newline \texttt{FINDPOLES\_POLYMP}(\texttt{findpoles\_polymp.f})}

These subroutines find the roots $t^P_{j}$, $j=1,2,...[n/2]$ of the polynomial $P_{n}(t)$ and  
the roots $t^Q_{j}$, $j=1,2,...[(n-1)/2]$ of the polynomial $Q_{n}(t)$,
\begin{eqnarray} \label{prod}
P_{n}(t)=p_{[n/2]} \prod_{j=1}^{[n/2]}(t-t^P_{j}), \quad
Q_{n}(t)=q_{[(n-1)/2]}\prod_{j=1}^{[(n-1)/2]}(t-t^Q_{j}).
\end{eqnarray}%
The  subroutines call internally subroutines  
\texttt{POLYROOTS} (for DP) or \texttt{POLYROOTS\_MP} (for MP),
which find roots of a complex polynomial using 
Weierstrass-Durand-Kerner-Dochev type algorithm  \cite{WDKD}.
Input and output for the subroutines
\texttt{FINDPOLES\_POLY} /\texttt{FINDPOLES\_POLYMP} is as follows,

\begin{small}
\begin{tabular}{l l l l}
\hline
input & zqn/zqnmp & DP/MP, complex & coefficients of $Q_{n}(t)$\\
output & zppole & DP, complex & poles after each iteration\\
output & zplf & DP, complex & poles after final iteration\\
output & zplcoff & DP, complex & coefficients $Q_n(t)$ after final iteration,\\
\hline
\end{tabular}
\end{small}

and similarly for
\texttt{FINDZEROS\_POLY} / \texttt{FINDZEROS\_POLYMP}:

\begin{small}
\begin{tabular}{l l l l}
\hline
input & zpn/zpnmp & DP/MP, complex & coefficients of $P_{n}(t)$\\
output & zpzero & DP, complex & poles after each iteration\\
output & zzef & DP, complex & poles after final iteration\\
output & zzcoff & DP, complex & coefficients $P_n(t)$ after final iteration.\\
\hline
\end{tabular}
\end{small}

\subsection{Subroutines \newline \texttt{FINDZEROS\_NAG} (\texttt{findzeros\_NAG.f}),
 \newline \texttt{FINDPOLES\_NAG} (\texttt{findpoles\_NAG.f})}

The function of these subroutines is the same as of \texttt{FINDPOLES\_POLY} and \texttt{FINDZEROS\_POLY}
but they use a rootfinder \texttt{CO2AFF} from 
the NAG Fortan Library \cite{NAGROOT} instead of \texttt{POLYROOTS}.
These were introduced in order to cross-check the accuracy of the roots found by \texttt{POLYROOTS}
in a DP calculation. With no access to the source code, we have been unable to convert NAG routines to multiple precision.

\subsection{Subroutine \texttt{SMOOTH} (\texttt{smooth.f})}

Poles and zeroes located in the vicinity of the real axis cause rapid variations of the 
phase of the fraction $P_{n}(t)/Q_{n}(t)$. Evaluation of the slowly changing potential phase
requires removal of such zeroes and poles from the ratio.
This is done in this subroutine in the following way: the user supplies a parameter $dxl$ defining
the width of a strip in the complex $t$-plane, $-dxl < Im t < dxl$, and contributions
from all poles and zeroes within the strip are removed from the products in Eqs.(\ref{prod}). 
The argument of the remainder is calculated as
\begin{equation}
 \phi(t) = Arg\Big[ \frac{\prod_{i}'(t-t^P_{i})}
                {\prod_{j}'(t-t^Q_{j})}\Big]
\end{equation} 
where primes indicate products restricted by the conditions $|Im t^P_{i}|>dxl$ and $|Im t^Q_{i}|>dxl$.
The inputs and outputs to this subroutine are as follows

\begin{small}
\begin{tabular}{l l l l}
\hline
input & ts & DP, real & shifted angular momentum or energy\\
input & zpzero(*) & DP, complex & zeros after each iteration\\
input & zppole(*) & DP, complex & poles after each iteration\\
output & tphs & DP, real & points on real axis \\
output & phs & DP, real & the phase $\phi(t)$.  \\
\hline
\end{tabular}
\end{small}

\subsection{Subroutine \texttt{FIT} (\texttt{fit.f})}

This subroutine fits the function $\phi(t)$ evaluated in  \texttt{SMOOTH}
to a quadratic form $\tilde{ \phi(t)} = afit(3)t^2+afit(2)t+afit(1)$ and 
determines the values of $afit(i)$, $i=1,2,3$.
Fitting is done by calling the routine \texttt{E02ACF} from the NAG Fortran Library
\cite{NAGFIT}.
The inputs and outputs to this subroutine are as follows:

\begin{small}
\begin{tabular}{l l l l}
\hline
input & EF & DP, real & values of $t$ \\
input & f & DP, real & $\phi (t)$ computed in  \texttt{SMOOTH}\\
input & NDAT=3 & integer & number of the coefficients \\
output & A(i), i=1,3 & DP, real & coefficients of the quadratic form\\
\hline
\end{tabular}
\end{small}

If the number of iterations is less then the value 'niter' specified by the user in the input file {\it param.pade}, $icheck < niter$,
the quadratic phase is removed from the input data in the main program; the calculation repeats  itself until $icheck = niter$ and the program proceeds to 
creating output. 



\subsection{Subroutine \texttt{SPADE} (\texttt{spade.f})}

This subroutine computes the \pd approximant, its phase, and the derivative of the phase
for the values of the argument between the  smallest and largest values specified by the 
user in  \texttt{OPEN\_FILES}. The number of the values is \texttt{nprnt} as specified in the input file
{\it param.pade}.   Sum of all the phases subtracted during iterations, 
preconditioning and parity are returned. The normalisation factor $K_N$ in Eqs. (\ref{7}) and
(\ref{9}) is computed from the values $\phi$'s in Eq.(\ref{frac}) \cite{PAD1}, calculated in the function
\texttt{ZFAC} (\texttt{zfac.f}) called internally from the coefficients of the continued fraction (\ref{frac}) \cite{PAD1}.

The inputs and outputs to this subroutine are as follows:

\begin{small}
\begin{tabular}{l l l l}
\hline
input & zphi & DP, complex & set of $\phi$'s in Eq.(\ref{frac}) \\
input & phi1, phi2, phi3 (*) & DP, real & preconditioning coefficients\\
input & xfit(i)(*) & DP, real & xfit(i)=$\sum_{all\quad iterations}$afit(i), i=1,3 \\
input & zzef (*) & DP, complex & zeros after final iteration\\
input & zplf(*) & DP, complex & poles after final iteration\\
output & zst & DP, complex & values for which the approximant is evaluated \\
output & zspade & DP, complex & values of the approximant\\
output & y(*) & DP, real & phase of the approximant\\
output & dydx(*) & DP, real & derivative of the phase of the approximant\\
\hline
\end{tabular}
\end{small}

\subsection{Subroutine \texttt{RESIDUE} (\texttt{residue.f})}

This subroutine computes the residues at the poles of a \pd approximant [cf. Eq.(\ref{8})]

The inputs and outputs to this subroutine are as follows:

\begin{small}
\begin{tabular}{l l l l}
\hline
input & phi1, phi2, phi3 (*) & DP, real & preconditioning coefficients\\
input & xfit(i)(*) & DP, real & xfit(i)=$\sum_{all\quad iterations}$afit(i), i=1,3 \\
input & zzef(*) & DP, complex & zeros of the approximant\\
input & zrt & DP, complex & poles of the approximant\\
output & zresid & DP, complex & set of residues for the poles of the approximant\\
\hline
\end{tabular}
\end{small}

\subsection{Subroutine \texttt{OUTPUT} (\texttt{output.f})}

This subroutine writes out the output files as follows.
If the short print out is produced, \texttt{ (Index(run\_option,"full\_print") = 0)},
the parameters of the \pd approximant (\ref{7}) or (\ref{9}) are written on {\it out\_pade} file
in the following order: 

* number of zeroes ($kp$), number of poles ($kq$),

* $kp$ zeroes positions, 

* $kq$ poles positions,

 * coefficients $a$, $b$ and $c$ of the quadratic phase,
 
* the constant factor $K_n$ 

* a flag which has the values $1(0)$ if the parity change
has (has not)  been applied during the calculation.

For a test calculation involving the initial data supplied with the program, 
\texttt{(Index(test\_option,"test") /= 0)} the same data is written on the file {\it out\_pade\_test}
but in a specific format chosen to avoid machine dependent differences which may occur in the last digits of the output.

Finally, if a long print out is produced \texttt{ (Index(run\_option,"full\_print") /= 0)} additional data is
written out as follows:

1. the real part and the absolute value of the original input data selected in \texttt{OPEN\_FILES} are written out on  ' {\it inputvals} file.

2. A grid of $nprint$ equidistant points is introduced in the region containing the input values ($J$ or $E$).
 Coordinates of the grid points and the corresponding values of the real part and the absolute value of the \pd approximant
 are written on  {\it smprod} file.
 
 3.  Coordinates of the same grid points, the corresponding values of the (continuous) phase of the \pd approximant  and the values of its derivative  are written on  the file {\it phdph}.
 
4. Real  and imaginary parts of all the zeroes of the \pd approximant are written on  the file' {\it zeros}

5. Real  and imaginary parts of all the poles of the \pd approximant are written on  the file {\it poles}

6. Real  and imaginary parts of all the poles and the real and imaginary parts of the corresponding residues are written on  the file {\it resids}

Upon a successful termination of a calculation a message  \texttt{'NORMAL JOB TERMINATION'}
is printed on the screen.
\section{Additional routines}
\subsection{Subroutine \texttt{VALIDATION}  (\texttt{validation.f})} This routine recalculates the values of the \pd approximant at the initial input points  and compares them with the input values of the $S$-matrix element. If validation is done, the routine writes onto \texttt{job.log} after the line
 {\it 'INPUT TEST'} the value 
$\sum_{J=J_{min}}^{J_{max}}|S(E,J)-S^{Pade}(E,J)|$ or $\sum_{j=j_{min}}^{j_{max}}|S(E_j,J)-S^{Pade}(E_j,J)|$ depending on whether CAM or CE poles are obtained.

\subsection{Subroutines  \texttt{TRANSMP90}(\texttt{transmp90.f})}
This is a package of FORTRAN 77 routines developed by D.H.Bailey \cite{MPFUN}, which works in conjunction with {\it MPFUN}. It translates a standard FORTRAN 77 code into a code calling \texttt{MPFUN} multiple precision routines.

\section{Installation}

Installing and testing {\it PADE\_II} involves unpacking the software and running the test suite.

\subsection{ System requirements}
This version of {\it PADE\_II} is intended for computers running the Linux/Unix operating system. 

Other requirements include:  FORTRAN  compilers,  $NAG$ Numerical Libraries and 
 a translator of FORTRAN codes to multiprecision  (included in the {\it PADE\_II} package)
\subsection{Unpacking the software}
The software is distributed in the form of a gzip'ed tar file which contains the {\it PADE\_II} source code and test suite, as well as the scripts needed for running and testing the code. 

To unpack the software, type the following command: 

\begin{verbatim}
tar -xzvf PADE_II.tgz
\end{verbatim}

This will create a top-level directory called PADE and subdirectories as shown in Fig 1.
\begin{figure}
\begin{center}
\includegraphics[angle=0,width=10 cm]{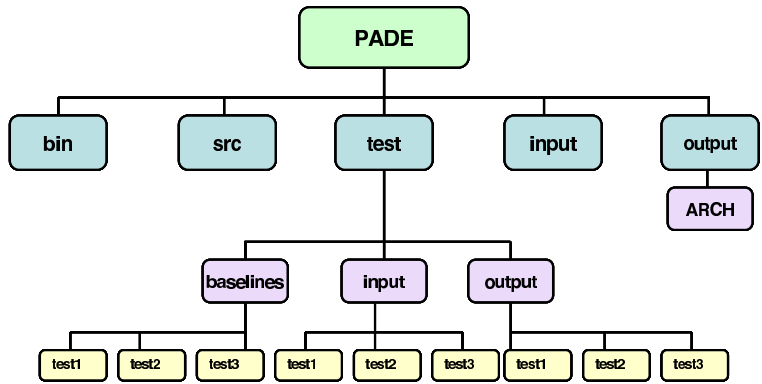}
\end{center}
\caption{Schematic structure of the package}
\end{figure}
For users' benefit we supply a file README in directory PADE. The file provides a brief summary of the code structure and basic instructions for its users. 
\subsection{Building {\it PADE\_II} executables from source}
To build the executables {\it PADE\_II} and the utilities {\it VALIDATE} and {\it TRANSMP90} (a FORTRAN -multiprecision translator) perform the following steps:
\begin{verbatim}
cd src
\end{verbatim}
Change the definition of \texttt{F77, FFLAGS, NAGPATH} and \texttt{LFLAGS} in \texttt{Makefile} if necessary.
\begin{verbatim}
make all
\end{verbatim}
This will create the binaries {\it PADE\_II}, {\it VALIDATE}, and {\it TRANSMP90} in directory 
\texttt{PADE/bin}.

To build each executable separately simply type 
\begin{verbatim}
make exe_name 
\end{verbatim}
where  {\bf exe\_name} is either {\it PADE\_II} or VALIDATE or TRANSMP.

\subsection{ Validation tests}

The input for three jobs,  {\it test1},  {\it test2} and  {\it test3}, are provided in directories {\it test/input/test\_name} where  {\it test\_name} is either  {\it test1} or  {\it test2} or  {\it test3}.
To submit and run a test suite, type the following commands:
\begin{verbatim}
cd PADE/test
\end{verbatim}
\begin{verbatim}
./run_TEST
\end{verbatim}
All tests are run in separate directories, {\it PADE/test/output/test1}, {\it PADE/test/output/test2} and {\it PADE/test/output/test3}.
Each test takes about from few seconds to several minutes to run on a reasonably modern computer.
To analyse the test results, inspect the message at the conclusion of the testing process. 
The message
\begin{verbatim}
Test test_name was successful
\end{verbatim}
confirms that the code passed the validation test {\it test\_name} . The results of the simulation can be viewed in the {\it output/test\_name} directory.  
The message
\begin{verbatim}
Your output differs from the baseline!
\end{verbatim}
means that the calculated data significantly differ from that in the baselines. Check the files output/test\_name/diff\_file to judge the differences. 

\section{Running {\it PADE\_II}}

  \subsection{Computational modules}
    
{\it PADE\_II}:    the code for calculating the poles and zeros, as well as the poles residues in the plane of complex angular momentum / complex energy using  \pd approximation of type II.

{\it VALIDATE}:   the code validating the results obtained using {\it PADE\_II}. 

{\it TRANSMP}:     a translation program allowing for extension of the FORTRAN 77 language to multiple precision data types.
	
Running {\it PADE\_II} involves the following steps:

1. Create the input data,

2. Specify the run options and execute {\it PADE\_II},

3. Validate the output data (optional).

  \subsection{ Creating input data}
Two input files are required for running calculations: a parameter file, {\it param.pade}, and the file containing the input data to be \pd approximated. The latter can have an arbitrary name.
Examples of input files can be found in directory {\it PADE/input}. 

 \subsection{Executing a simulation}
The script {\it run\_PADE} in {\it PADE/ directory} automates calculations. The following assumptions are made in the scripts: 

* all  binaries are placed in {\it PADE/bin}

* input files are located in directory {\it PADE/input} 

* output files can be found in {\it PADE/output} on completion of the calculation.
 
Four options can be specified in the script {\it run\_PADE} before running: 

{\it inputfile: }  the name of the input file. Default is {\it input}. The user has freedom in choosing a name for the input file

{\it runoption: }                an option controlling a length of output. Please set {\it runoption} to {\it production} for a production run or to {\it full\_print}  for the detailed output (see section 6.13 for further detail). The default is {\it production}

 {\it testoption: }   an option allowing for a use of the test script {\it run\_TEST} in {\it PADE/test} directory. Please set {\it testoption} to {\it test} for testing or leave blank otherwise. Default is a blank value.

{\it validation: } an option allowing for validating the output data using the utility validate. Please set {\it validation} to {\it  valid\_yes} to validate the results or to {\it valid\_no} otherwise. The default is {\it valid\_yes}. We encourage users to choose the default option. 

We recommend running a calculation in directory {\it PADE/. }The command
\begin{verbatim}                    
../bin/run_PADE
 \end{verbatim}
immediately starts the calculation.   
   \subsection{Output data}
On completion of the calculation all output files will be redirected to {\it PADE/output}. The directory {\it ARCH} will be created automatically in {\it PADE/output} if it does not exist yet.  
Please take care of the output from the previous run in directory {\it PADE/output} as they will be automatically removed at the start of the next calculation. It does not apply to the content of directory {\it ARCH}.
The following output files can be found on completion of the {\it PADE\_II} run with the chosen {\it runoption} to be {\it  production}:

      {\it out\_pade}  contains all significant calculated data, such as zeros, poles, phase coefficients 
      
       {\it out\_pade\_test}   contains the same data as in  {\it out\_pade} but in the different format: with 8 digits shown after the decimal point 
       
       {\it resides}    contains the calculated resides   
       
       {\it job.log}    contains  {\it PADE\_II} job statistics file and the summary of the validation procedure (if any)
       
     {\it summary.\$DATE}   stores the name of the input file used.  {\it\$DATE} contains origination date and time. The file is located in {\it PADE/output/ARCH}.
     
If {\it runoption} is chosen to be {\it full\_print} then a number (depending on parameters used) of output files will be created in directory {\it PADE/output}. Those files provide detailed information and extra control of the calculation (see 6.13).


\newpage
\hspace{1pc}
\section {Test runs}
\bigskip

Three suitable test runs of the $\texttt{PADE\_II}$ program package are provided. 
The input and output files for these tests are included in the package.

\subsection{Test 1: The hard sphere model (Regge poles)}
The first test 
involves $S$ matrix element  for potential (single channel) scattering off a hard sphere of a radius $R-d$ surrounded by a thin semi-transparent layer of a radius $R$, 
so that the spherically symmetric potential $V(r)$ is infinite for $r<R-d$ and $\Omega \delta(r-R)$ ($\delta(x)$ is the Dirac delta) elsewhere. The energy of a non-relativistic particle is $E=k^2/2$, 
where we have put to unity $\hbar$ as well as the particle's mass.
A detailed discussion of this model can be found in Refs.\cite{G1} and \cite{ICS3}.
The input file $smre.test1$ contains $40$ partial waves, $J=0,1...39$,
for a model with $d/R=3.5/20$, $R\Omega=50$ and $kR=20$. 
The input files  {\it param.pade} and {\it input\_file} are as follows
\begin{verbatim}
ipar  iprec  imult   nstime   nprint     fac
 0      1       0       1       2000   0.000001
\end{verbatim}
and
\begin{verbatim}
        nread     niterr       shift   jstart   jfin   inv  dxl 
          40           2          20       1     39    -1   1.5
  0.000000000000000E+000 -0.453586393436287      -0.891212311230866
   1.00000000000000       0.543023350525081       0.839717595852626
   2.00000000000000      -0.704041319711519      -0.710159010460940
   3.00000000000000       0.888058065387598       0.459731304677021
   4.00000000000000      -0.998289775531788      -5.845959347012353E-002
   5.00000000000000       0.890738274698926      -0.454516584940946
   6.00000000000000      -0.434473236117557       0.900684743457741
   7.00000000000000      -0.333662099368660      -0.942692740740534
   8.00000000000000       0.977155982861081       0.212523375558536
   9.00000000000000       0.720595526411046       0.693355671583052
   10.0000000000000       0.664031500188029      -0.747704598593613
   11.0000000000000       0.666737175185565       0.745292921760011
   12.0000000000000      -0.856249634342491       0.516562255384915
   13.0000000000000      -0.425006254859363      -0.905190412747737
   14.0000000000000       0.899356824643227      -0.437215395391851
   15.0000000000000       0.564958024126621       0.825119646460405
   16.0000000000000      -0.624574779169686       0.780965008963358
   17.0000000000000      -0.976521727561331      -0.215418930460241
   18.0000000000000      -0.404573635897627      -0.914505425427632
   19.0000000000000       0.324581052098112      -0.945857886058409
   20.0000000000000       0.770295446722293      -0.637687168413246
   21.0000000000000       0.944712675162273      -0.327899315930580
   22.0000000000000       0.990708814056840      -0.136000168198755
   23.0000000000000       0.998911104661369      -4.665409932903946E-002
   24.0000000000000       0.999908936584357      -1.349513018612076E-002
   25.0000000000000       0.999994360792535      -3.358330407996912E-003
   26.0000000000000       0.999999731959201      -7.321758842783459E-004
   27.0000000000000       0.999999989939087      -1.418514237449860E-004
   28.0000000000000       0.999999999695326      -2.468496879744044E-005
   29.0000000000000       0.999999999992432      -3.890541681219141E-006
   30.0000000000000       0.999999999999844      -5.590812250283133E-007
   31.0000000000000       0.999999999999997      -7.367561396623236E-008
   32.0000000000000        1.00000000000000      -8.958077930733973E-009
   33.0000000000000        1.00000000000000      -1.020177826822646E-009
   34.0000000000000        1.00000000000000      -1.241835300295753E-010
   35.0000000000000        1.00000000000000       1.384853577811148E-011
   36.0000000000000        1.00000000000000       7.957477362324928E-014
   37.0000000000000        1.00000000000000       1.036005717080183E-014
   38.0000000000000        1.00000000000000       9.917113494880074E-015
   39.0000000000000        1.00000000000000       1.000697917812276E-014
\end{verbatim}								   
respectively. The output file 	{\it out\_pade} contains the data which define a 
$[19/19]$ \pd approximant (cf.Eq.\ref{7})

\begin{verbatim}
          19          19
 ZEROES
           1 (39.4911990444339,6.19497800757232)
           2 (36.9799732414393,8.43963299665421)
           3 (34.0133580781357,10.7738603477704)
           4 (18.1696986797484,18.1210578980068)
           5 (12.2668387160508,19.4124385241265)
           6 (5.83305872468270,20.3995836982532)
           7 (-3.02720086016028,20.1000359128290)
           8 (24.6628733158064,2.67477729890608)
           9 (-12.0363094250139,11.9513828909703)
          10 (-12.1776119341912,1.25316966687318)
          11 (9.14176939602146,-0.234322402585798)
          12 (14.5523482093869,-1.109430235203338E-003)
          13 (21.8125711139416,-4.39732130205730)
          14 (25.1197847383941,-10.5991913301718)
          15 (26.2269224845999,-12.0435440258052)
          16 (23.6585655155745,-7.73331773608408)
          17 (24.6544987721333,-2.67109330547254)
          18 (35.7440073905768,-3.138594025574343E-003)
          19 (41.8597046798259,3.61274050690063)
 POLES
           1 (35.7440073903667,-3.138542834584743E-003)
           2 (24.6628733369383,2.67477729572751)
           3 (23.6585655606588,7.73331699080323)
           4 (26.2268330879602,12.0434263955026)
           5 (25.1197975398861,10.5992368004622)
           6 (21.8125711134792,4.39732130283915)
           7 (14.5523482093869,-1.109430235909550E-003)
           8 (9.14176939602135,0.234322402586938)
           9 (-12.1774051901505,-1.25342226018641)
          10 (-12.0359449939580,-11.9511561142494)
          11 (24.6544987511749,-2.67109330818693)
          12 (-3.02710863601014,-20.0995263248157)
          13 (5.83307474171240,-20.3992358761048)
          14 (12.2668862760970,-19.4121731020176)
          15 (18.1697778070471,-18.1208548568950)
          16 (34.0131260419783,-10.7739117218385)
          17 (36.9796907970374,-8.43956124986996)
          18 (39.4909636463193,-6.19477828169534)
          19 (41.8595888739847,-3.61248175364997)
 PHASE COEFF a+b*J+c*J^2
   2.67555460298088        2.73887621121985      -4.727964256522786E-002
 CONST FACTOR K_n
 (0.357477628915536,0.933794075548949)
 PARITY CHANGE FROM ORIGINAL DATA
           0
\end{verbatim}

Figure 2 shows the behaviour of the  \pd approximant along the real $J$ axis.
The pole/zero configuration of the $[19/19]$ \pd approximant is shown in Fig. 3a.

\begin{figure}\label{shit}
\begin{center}
\includegraphics[angle=-90,width=10 cm]{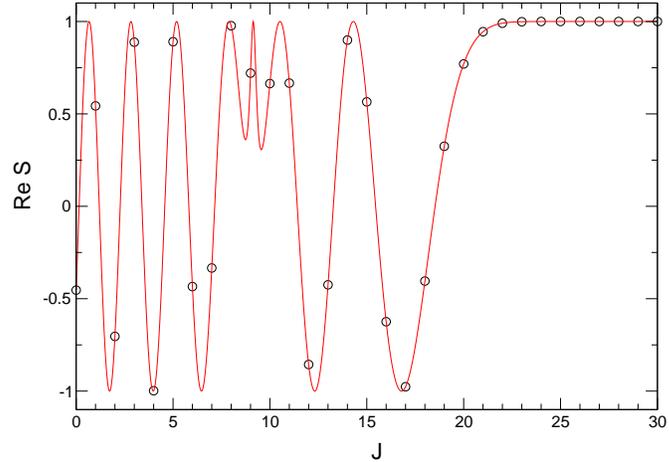}
\end{center}
\caption{Real parts of the input values of the hard-sphere model matrix element
$S(E,J)$, $J=0,1,...39$ (circles) and the real part of the $[19/19]$ \pd approximant constructed from these
values (solid).}
\end{figure}
Figure 3b shows the results of two similar calculations ($nstime=2$ chosen in the input {\it param.pade} )  .  However, in
each calculation a random noise with the noise factor in Eq.(\ref{noise}) $fac=10^{-6}$ has been 
added to the input data. As discussed in Sect.2, additional contamination with a non-analytical noise
results in squeezing of the domain of validity of an approximant as well as in production of additional Froissart doubles close to the real axis.
\newpage
\begin{figure}\label{1F}
\begin{center}
\includegraphics[angle=-0,width = 12 cm]{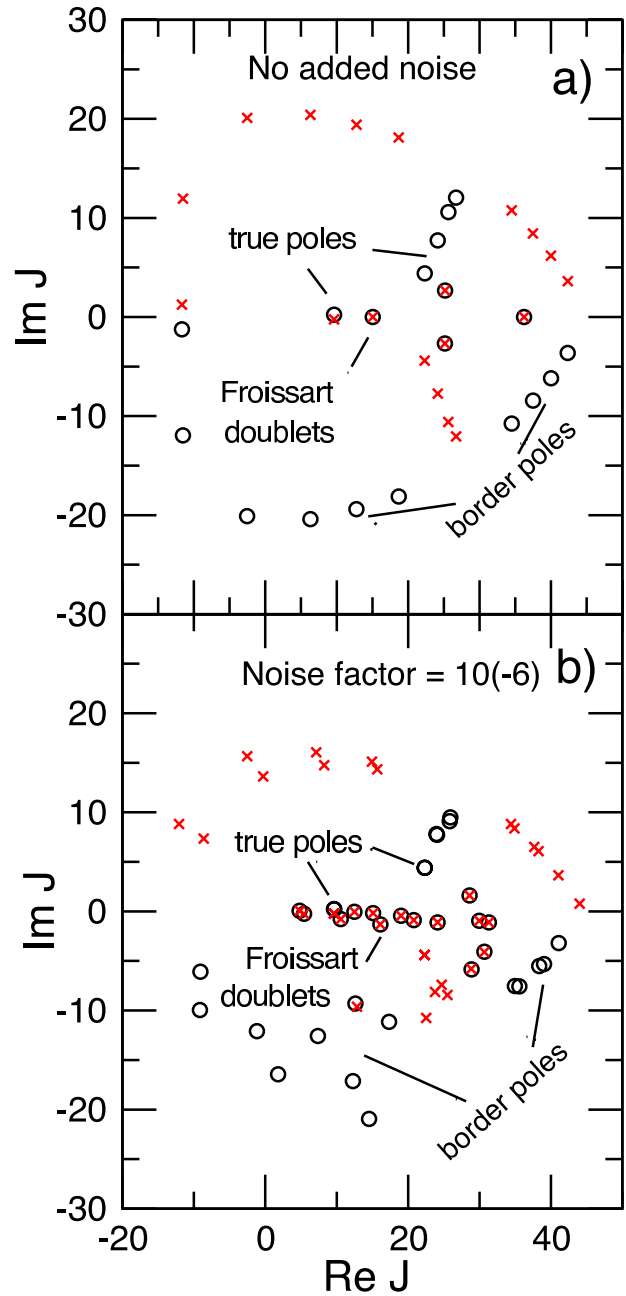}
\end{center}
\caption{a) Positions of poles (circles) and zeros (crosses) in the complex angular momentum plane for 
the hard-sphere model. Three types of poles are indicated. The true poles include an
isolated resonance with $Re J \approx 10$ on a metastable state trapped between the core and the outer layer as
well as the first four poles of the infinite sequence corresponding to diffraction on the outer layer
\cite{G1}. There are also several Froissart doublets. Remaining poles and zeroes mark a boundary  beyond which the approximant fails,
b) Poles and zeroes of two $[19/19]$ \pd approximants constructed from the same input values,
but contaminated with random noise with $fac=10^{-6}$. }
\end{figure}
\newpage
\subsection{Test 2: The hard sphere model (complex energy poles)}
For the second test the input file $smrj.test2$ contains $49$ values 
of $S(E,J=10)$ on the equidistant energy grid $200 \le R^2E_j=R^2 k_j^2/2 < 250$
for the same model with $d/R=3.5/20$, $R\Omega=50$. 
The input files  {\it param.pade} and {\it input\_file} are as follows
\begin{verbatim}
ipar  iprec  imult    nstime  nprint      fac
 0      0       0       0      2000   0.000001
\end{verbatim}
and
\begin{verbatim}
nread niter    shift    jstart     jfin    inv dxl 
  50     5       225.       1        49    -1  1.5 
   200.000000000000       0.664031500188029      -0.747704598593613
   201.000000000000       0.640154019051953      -0.768246595756617
   202.000000000000       0.624758751680046      -0.780817841880672
   203.000000000000       0.621472411367862      -0.783436048384688
   204.000000000000       0.635123092357332      -0.772410938266969
   205.000000000000       0.671965187035159      -0.740582735022093
   206.000000000000       0.739032006430064      -0.673670315118571
   207.000000000000       0.839572088290516      -0.543248293659080
   208.000000000000       0.954602056895436      -0.297884059612129
   209.000000000000       0.992695675238598       0.120645332949871
   210.000000000000       0.761692689904593       0.647938458610000
   211.000000000000       0.223560893654405       0.974689964464827
   212.000000000000      -0.304719743630442       0.952442060096988
   213.000000000000      -0.627460687663612       0.778648242428316
   214.000000000000      -0.789516598726692       0.613729207660053
   215.000000000000      -0.866908960431839       0.498466502709047
   216.000000000000      -0.903313320122455       0.428981404829331
   217.000000000000      -0.919143298112510       0.393923339667577
   218.000000000000      -0.923506334424972       0.383583172567504
   219.000000000000      -0.920433057855924       0.390900225129116
   220.000000000000      -0.911688160524598       0.410882827530277
   221.000000000000      -0.898018594051828       0.439957503330923
   222.000000000000      -0.879718432979564       0.475494982808420
   223.000000000000      -0.856891241732639       0.515497235532739
   224.000000000000      -0.829574633076335       0.558395852560049
   225.000000000000      -0.797800024420750       0.602922151719648
   226.000000000000      -0.761621008213086       0.648022715534326
   227.000000000000      -0.721126129965770       0.692803799556982
   228.000000000000      -0.676443760254596       0.736494290006801:
   229.000000000000      -0.627742897353397       0.778420744085332
   230.000000000000      -0.575231856010094       0.817990410598549
   231.000000000000      -0.519155855614356       0.854679587670915
   232.000000000000      -0.459794039256912       0.888025586041198
   233.000000000000      -0.397456205017629       0.917621144641396
   234.000000000000      -0.332479399617643       0.943110517823808
   235.000000000000      -0.265224455432244       0.964186697813587
   236.000000000000      -0.196072515839709       0.980589398541656
   237.000000000000      -0.125421575098095       0.992103537187482
   238.000000000000      -5.368304983650439E-002  0.998558025434802
   239.000000000000       1.872160509145225E-002  0.999824735392559
   240.000000000000       9.136422532965484E-002  0.995817542690383
   241.000000000000       0.163813199773938       0.986491376333227
   242.000000000000       0.235636656616718       0.971841224716515
   243.000000000000       0.306405561742876       0.951901061945533
   244.000000000000       0.375696718367498       0.926742669681230
   245.000000000000       0.443095656457025       0.896474338299161
   246.000000000000       0.508199400868439       0.861239437646094
   247.000000000000       0.570619106928023       0.821214853012697
   248.000000000000       0.629982552997152       0.776609285882670
   249.000000000000       0.685936479728164       0.727661422488601
   250.000000000000       0.738148766653856       0.674637975722825
   10.5000000000000        20.0000000000000        3.50000000000000
   5.00000000000000       0.000000000000000E+000
   \end{verbatim}		   
respectively. The output file {\it out\_pade} contains the data which defines a 
$[24/24]$ \pd approximant (cf.Eq.\ref{9})

\begin{verbatim}
          24          24
 ZEROES
           1 (268.545717720799,11.4671502976118)
           2 (261.728173386068,19.6753217953572)
           3 (254.557738718419,26.5701017472017)
           4 (245.881831513906,33.4912944344641)
           5 (232.951306941574,1.877662435186347E-004)
           6 (224.121052772429,4.19191611872841)
           7 (220.411919658935,1.15434928347258)
           8 (217.652385494978,2.852399491808721E-002)
           9 (210.331899212664,2.72550066138290)
          10 (209.654815139940,2.641960499678558E-004)
          11 (212.868385932177,3.598159766224907E-004)
          12 (186.524319063303,-9.14781503752908)
          13 (206.326496566314,1.959778648327130E-004)
          14 (192.395056901391,-15.3604166393560)
          15 (198.447273067597,-20.3445869086093)
          16 (205.051304571580,-24.9792759962048)
          17 (212.514236963499,-29.6992072722432)
          18 (221.649194114141,-34.9881538867756)
          19 (220.510303177377,-1.14210693606552)
          20 (224.031598890991,-4.30241298271368)
          21 (226.023976865667,2.261727493434314E-002)
          22 (238.490995343524,-1.47246931619824)
          23 (242.730754962705,-3.477317333857598E-004)
          24 (238.491169699397,1.46793956930393)
 POLES
           1 (242.730754962705,-3.477317319122845E-004)
           2 (238.491169699401,1.46793956929681)
           3 (226.023976865667,2.261727493435679E-002)
           4 (224.121052772959,4.19191611897241)
           5 (220.411919658902,1.15434928347629)
           6 (221.648098623382,34.9885968283023)
           7 (212.513501359997,29.7001476182166)
           8 (205.051076178872,24.9802350950195)
           9 (198.447298402548,20.3453065442821)
          10 (192.395118933229,15.3609323651784)
          11 (206.326496566310,1.959780792806160E-004)
          12 (186.524346342521,9.14822455412355)
          13 (212.868385932166,3.598159852993832E-004)
          14 (209.654815139977,2.641956005816340E-004)
          15 (210.331899212667,-2.72550066136285)
          16 (217.652385494976,2.852399491882152E-002)
          17 (220.510303177407,-1.14210693605598)
          18 (224.031598890383,-4.30241298239934)
          19 (232.951306941574,1.877662434131332E-004)
          20 (245.882621687043,-33.4918893250634)
          21 (254.558185226003,-26.5708559777491)
          22 (261.728380822961,-19.6760176672578)
          23 (268.545803872059,-11.4677538269500)
          24 (238.490995343520,-1.47246931620535)
 PHASE COEFF a+b*J+c*J^2
  -283.165777224028        2.47637834572351      -5.407538683988272E-003
 CONST FACTOR K_n
 (0.908740385440017,0.417744816017221)
 PARITY CHANGE FROM ORIGINAL DATA
           0
 \end{verbatim}          
Figure 4 shows the behaviour of the  \pd approximant along the real $E$-axis.
The pole/zero configuration of the $[24/24]$ \pd approximant is shown in Fig. 5.
\begin{figure}
\begin{center}
\includegraphics[angle=-90,width=10 cm]{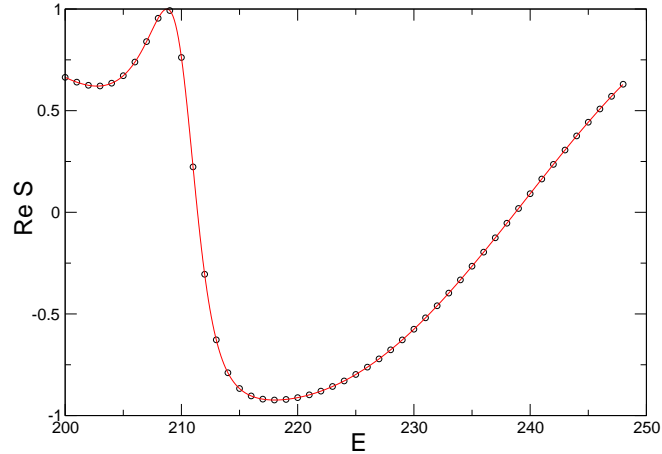}
\end{center}
\caption{Real parts of $S(E_m,J=10)$, $m=1,...49$ (circles) and the $[24/24]$ \pd approximant constructed from these
values (solid).}
\end{figure}
\begin{figure}
\begin{center}
\includegraphics[angle=-90,width = 12 cm]{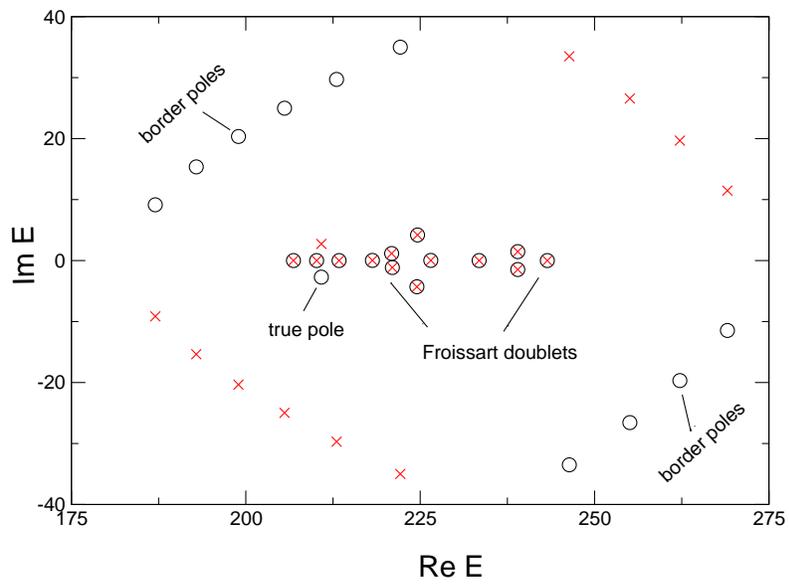}
\end{center}
\caption{Positions of poles (circles) and zeros (crosses) in the complex energy plane for 
10-th partial wave ($J=10$) of  the hard-sphere model. Three types of poles are indicated. The true pole with $Re E \approx 212$ corresponds to a metastable state trapped between the core and the outer layer. There are also several Froissart doublets. Remaining poles and zeroes mark a boundary  beyond which the approximant fails.}
\end{figure}
%
\subsection{Test 3: Multiple precision (Regge poles)}
The third test, whose purpose is to illustrate the use of the multiple precision option,
involves constructing a \pd approximant for the inelastic matrix element 
for the $He^+ - Ne$ inelastic collision model of Olson and Smith \cite{OLSON} computed by K.Thylwe \cite{KET}. The input consists of $111$ partial 
waves, i.e., the values of $S_{01}(E,J)$, $J=0,1, ...,110$ at a fixed value of the collision energy.
The input file  {\it param.pade} is 
\begin{verbatim}
ipar  iprec  imult    nstime  nprint      fac
 0      1       1      0        2000   0.000001
\end{verbatim}
and the input  {\it input\_file} together with the output  {\it out\_pade}, too long to be printed here, are supplied with the code. The number of input data is such that a double precision calculation 
(choosing 'imult'$=0$ in {\it param.pade}) produces incorrect \pd approximant which fails to reproduce the input values (Fig.6a). A calculation with a multiple precision of $60$ digits 
yields the correct approximant (Fig.6b). In general, it is recommended to repeat a DP calculation
with a choice of a higher precision, in order to see if this changes the computed approximant in 
any significant  way. 
\begin{figure}
\begin{center}
\includegraphics[angle=-90,width=12 cm]{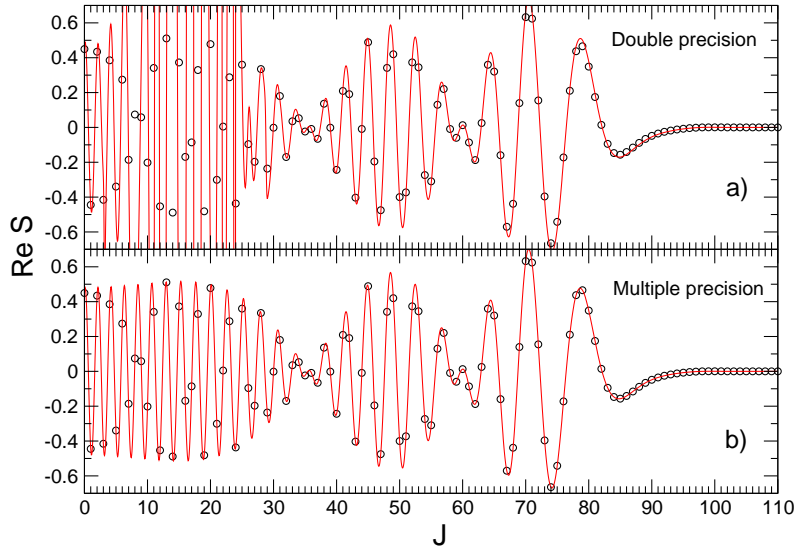}
\end{center}
\caption{Real parts of $S(E_m,J=10)$, $m=1,...49$ (circles) and the $[55/55]$ \pd approximant constructed from these
values (solid).}
\end{figure}
\section*{Acknowledgement}
This work was supported by the U.K. EPSRC through a grant GR/S03799/01 to CCP6 (Collaborative Computational Project No. 6 on Molecular Quantum Dynamics) for which this was the Flagship Project. DS is grateful to Daniel Bessis for useful discussions.


\begin{thebibliography}{99}
\bibitem{Skod} S. A. Harich, D. X. Dai, C. C. Wang, X. Yang, S. D. Chao and R. T. Skodje, Forward scattering due to slow-down of the intermediate in the H+HD$\leftarrow$D+H2 reaction, Nature, 419,  (2002) 281.

\bibitem{CLARY98} D. C. Clary, Quantum Theory of Reaction Dynamics,
Science, {\bf 279} (1998) 1879.

\bibitem{SKOU99} D. Skouteris, D. E. Manolopoulos, W. Bian, H-J Werner, 
L-H Lai and K. Liu, Van der Waals Interactions in the Cl + HD Reaction,
Science, {\bf 286} (1999) 1713.

\bibitem{CASA99} P. Casavecchia, N Balucani and G. G. Volpi, 
Cross beam studies of reaction dynamics,
Ann. Rev. Phys. Chem., {\bf 50} (1999) 347.

\bibitem{MILLER74} W. H. Miller, Adv. Chem. Phys., 
Classical-Limit Quantum Mechanics and the Theory of Molecular Collisions,
{\bf 25} 69 (1974) 69.

\bibitem{SKOU00} D. Skouteris, J. F. Castillo and D. E. Manolopoulos,
 ABC: a quantum reactive scattering problem, Comp. Phys. Commun,
{\bf 133} (2000) 128.

\bibitem{SCA02} S. C. Althorpe, F. Fern\'{a}ndez-Alonso, B. D. Bean, 
J. D. Ayers, A. E. Pomerantz, R. N. Zare and E. Wrede, 
Observation and interpretation of a time-delayed mechanism in hydrogen 
exchange reaction, Nature, {\bf 416} (2002) 67.

\bibitem{DARIO98} V. Aquilanti, S. Cavalli and D. De Fazio, 
Hyperquantization algorithm. I. Theory for triatomic systems,
J. Chem. Phys., {\bf 109} (1998) 3792.

\bibitem{DARIO02} V. Aquilanti, S. Cavalli, D. De Fazio, A. Volpi, 
A. Aguilar, X. Gim\'{e}nez and J. F. Lucas, 
Exact reaction dynamics by the hyperquantization algorithm: integral 
and differential cross sections for F + H$_2$, including long-range 
and spin-orbit effects, 
Phys. Chem. Chem. Phys., {\bf 4}, 401 (2002).
 \bibitem{FOOT} It is worth noting that the applications usually require the knowledge of
 only a few leading true pole positions and residues. The remaining poles and zeroes
 of the approximant are then used to reproduce the direct part of the $S$-matrix element on the real axis, which is independent of their individual positions and residues.
 

\bibitem{CON0}J.N.L. Connor,
 Molecular collisions and the semiclassical approximation,  Meldola Medal Lecture,
  Chemical Society Reviews,{\bf 5}, (1976), 125.

\bibitem{CON1} J.N.L. Connor, W. Jakubetz and C.V. Sukumar,
 Exact quantum and semiclassical calculation of the positions and residues of Regge poles for interatomic potentials,
  J. Phys. B, {\bf 9}, (1976), 1783.

\bibitem{CON2}J.N.L. Connor and W. Jakubetz,
 Rainbow scattering in atomic collisions: A Regge pole analysis, 
 Mol. Phys., {\bf 35}, (1978) 949.

\bibitem{CON3} J.N.L. Connor, D.C. Mackay and C.V. Sukumar, 
Quantum and semiclassical calculation of Regge pole positions and residues for complex optical potentials,
  J. Phys. B, {\bf 12}, (1979), L5151.

\bibitem{CON4} J.N.L.Connor,
 Semiclassical theory of elastic scattering,
 in  "Semiclassical methods in molecular scattering and spectroscopy." Proceedings of the NATO Advanced Study Institute held in Cambridge, England, in September, 1979. Edited by M.S. Child. Reidel, Dordrecht, The Netherlands, (1980) 45.

\bibitem{CON5} J.N.L. Connor, D. Farrelly and D.C. Mackay,
 Complex angular momentum analysis of diffraction scattering in atomic collisions,
  J. Chem. Phys., {\bf 74}, (1981) 3278.

\bibitem{CON6} K-E. Thylwe and J.N.L. Connor,
 A complex angular momentum theory of modified Coulomb scattering,
 J. Phys. A, {\bf 18}, (1985), 2957.

\bibitem{CON7} J.N.L. Connor, D.C. Mackay and K-E. Thylwe,
 Computational study and complex angular momentum analysis of elastic scattering for complex optical potentials,
  J.Chem. Phys., {\bf 85}, (1986) 6368.

\bibitem{CON8} J.N.L. Connor and K-E. Thylwe,
 Theory of large angle elastic differential cross sections for complex optical potentials: Semiclassical calculations using partial waves, l-windows, saddles and poles,
 J.Chem. Phys., {\bf 86}, (1987) 188.

\bibitem{CON9} J.N.L.Connor,
 New theoretical methods for molecular collisions: The complex angular momentum approach,  J. Chem. Soc., Faraday Transactions, {\bf 86}, (1990) 1627.

\bibitem{CON10} P. McCabe, J.N.L. Connor and K-E. Thylwe,
 Complex angular momentum theory of molecular collisions: New phase rules for rotationally inelastic diffraction scattering in atom homonuclear-molecule collisions,
  J.Chem. Phys., {\bf 98}, (1993) 2947.

\bibitem{Brink} D.M. Brink, {\it Semi-classical Methods in Nucleus-Nucleus Scattering},
Cambridge University Press, Cambridge, 1985.
\bibitem{G1} D. Sokolovski and  A. Z. Msezane, 
Semiclassical complex angular momentum theory and \pd reconstruction 
for resonances, rainbows, and reaction thresholds, 
Phys. Rev. A., {\bf 70} (2004) 032710.

\bibitem{G2} D. Vrinceanu, A. Z. Msezane, D. Bessis, J. N. L. Connor 
and  D. Sokolovski, Chem. Phys. Lett., \pd reconstruction of Regge poles 
from scattering matrix data for chemical reactions,
{\bf 324} (2000) 311.

\bibitem{G3} D. Sokolovski, S. Sen, 
On the type II \pd reconstruction of a scattering matrix element,
\textit {Semiclassical and other Methods for Understanding Molecular 
Collisions and Chemical reactions},
Collaborative Computational Project on Molecular Quantum 
Dynamics (CCP6), Daresbury, UK, (2005) 104.

\bibitem{HCL1} D. Sokolovski, J. N. L. Connor and G. C. Schatz, 
New uniform semiclassical theory of resonance angular scattering for 
reactive molecular collisions, Chem. Phys. Lett., {\bf 238} (1995) 127.

\bibitem{HCL2} D. Sokolovski, J. N. L. Connor and G. C. Schatz,
Complex angular momentum analysis of resonance scattering in the 
Cl + HCl $\rightarrow$ ClH + Cl reaction, J. Chem. Phys., {\bf 103} 
(1995) 5979.


\bibitem{FH1} D. Sokolovski, J. F. Castillo and C. Tully,
Semiclassical angular scattering in the F + H$_2$ $\rightarrow$ HF + H 
reaction: Regge pole analysis using the \pd approximation, Chem. Phys. 
Lett., {\bf 313} (1999) 225.

\bibitem{FH2} D. Sokolovski, J. F. Castillo, Differential cross 
sections and Regge trajectories for the F + H$_2$ $\rightarrow$ HF + H 
reaction, {\bf 2} (2000) 507.

\bibitem{FH3}D. Sokolovski, S.K.Sen, V.Aquilanti, S.Cavalli and D.De Fazio, 
Interacting resonances in the  F + H$_2$ reaction revisited: Complex terms, Riemann 
surfaces, and angular distributions 
J. Chem. Phys., {\bf 126} (2007) 084305.

\bibitem{FH4}D. Sokolovski, D.De Fazio, S.Cavalli and V.Aquilanti,
Overlapping resonances and Regge oscillations in the state-to-state integral 
cross sections of the  F + H$_2$ reaction, 
J. Chem. Phys., {\bf 126} (2007) 12110.

\bibitem{FH5}D. Sokolovski, D.De Fazio, S.Cavalli and V.Aquilanti,
On the origin of the forward peak and backward oscillations in the the  F + H$_2$(v=0)
$\rightarrow$ HF(v'=2) + H  reaction, 
Phys.Chem.Chem.Phys., {\bf 9} (2007) 1.



\bibitem{HD1} F. J. Aoiz, L. Ba\~{n}ares, J. F. Castillo and 
D. Sokolovski, Energy dependence of forward scattering in the 
differential cross section of the H + D$_2$ $\rightarrow$ 
HD($v^\prime = 3, j^\prime = 0$) + D reaction, 
J. Chem. Phys., {\bf 117} (2002) 2546.

\bibitem{HD2} D. Sokolovski, Glory and thresholds effects in 
H + D$_2$ reactive angular scattering, Chem. Phys. Lett., 
{\bf 370} (2003) 805.


\bibitem{FHD1} D. Sokolovski, Complex-angular-momentum analysis of 
atom-diatom angular scattering: Zeros and poles of the S matrix,
Phys. Rev. A., {\bf 62} (2000) 024702-01.

\bibitem{IH1} D. Sokolovski, A.Z. Msezane, Z. Felfli, S.Yu. Ovchinnikov and J.H. Macek,
What can one do with Regge poles?,
Nuclear Instruments and Methods in Physics Research Section B: Beam Interactions with Materials and Atoms, Volume {\bf 261}, (2007)133.

\bibitem{IH2} A.J. Totenhofer, C. Noli and J.N.L. Connor,
 Dynamics of the I + HI $\rightarrow$ IH + I reaction: Application of nearside-farside, local angular momentum and resummation theories using the Fuller and Hatchell decompositions,
 Physical Chemistry Chemical Physics, {\bf 12},(2010), in print.

\bibitem{ICS1} J. H. Macek, P. S. Krstic, and S. Yu. Ovchinnikov, 
Regge Oscillations in Integral Cross Sections for Proton Impact on Atomic Hydrogen,
Phys. Rev. Lett. {\bf 93},  (2004) 183203.

\bibitem{ICS2}D. Sokolovski, D.De Fazio, S.Cavalli and V.Aquilanti,
Overlapping resonances and Regge oscillations in the state-to-state integral 
cross sections of the  F + H$_2$ reaction, 
J. Chem. Phys., {\bf 126} (2007) 121101.

\bibitem{ICS3} D. Sokolovski, Complex-angular-momentum  (CAM) 
route to reactive scattering resonances: from a simple model to the F + H$_2$ $\rightarrow$ 
HF + H reaction ,
Phys. Scr., {\bf 78} (2008) 058118.

\bibitem{SE1} P. G. Burke and C. Tate, 
A program for calculating Regge trajectories in potential scattering,
Comput. Phys. Commun. {\bf 1}, 
1969, 97.
\bibitem{SE2} 
D. Sokolovski, Z. Felfli, S. Yu. Ovchinnikov, J. H. Macek, and A. Z. Msezane,
 Regge oscillations in electron-atom elastic cross sections, 
Phys. Rev. A {\bf 76},(2007)  012705.
\bibitem{PAD1} D. Bessis, A. Haffad, and A. Z. Msezane,
Momentum-transfer dispersion relations for electron-atom cross sections,
Phys. Rev. A., {\bf 49} (1994) 3366.

\bibitem{PAD2} G. A. Baker, Jr., \textit {The essentials of \pd 
Approximations}, Academic, New York, 1975.
\bibitem{WDKD} M.S. Petkovic,  C.Carstensen. and M. Trajkovic,  
Weierstrass formula and zero-finding methods, Numerische Mathematik, {\bf 69}: (1995).

\bibitem{MPFUN} D. H. Bailey, Algorithm 719, "Multiprecision translation 
and execution of Fortran programs". ACM Transactions on Mathematical 
Software, 19(3):288 (1993).

\bibitem{DSJNLCPL99} S. Sokolovski, J. N. L. Connor, 
Semoclassical nearside-farside theory for inelastic and reactive 
atom-diatom collisions,
Chem. Phys. Lett., {\bf 305} (1999) 238.

\bibitem{GAUTSCHI} W. Gautschi, in \textit {Handbook of Mathematical 
Functions}, edited by M. Abramowitz and I. A. Stegun (Harri Deutsch,
Htun, 1984).

\bibitem{FORD90} J. N. L. Connor,
J. Chem. Soc. Faraday Trans., {\bf 305} (1990) 1627.

\bibitem{NAGNOISE} Numerical Algorithms Group, Fortran Library 
Manual, Mark 19, subroutine G05CAF (NAG, OXFORD, 2002).

\bibitem{NAGROOT} Numerical Algorithms Group, Fortran Library
Manual, Mark 21, subroutine C02AFF (NAG, OXFORD, 2004).

\bibitem{NAGFIT} Numerical Algorithms Group, Fortran Library
Manual, Mark 21, subroutine E02ACF (NAG, OXFORD, 2004).

\bibitem{OLSON} R.E. Olson and F.T.Smith  Phys. Rev. A. {\bf 3}, (1971) 1607;  Erratum, Phys. Rev. A. {\bf 6},  (1972) 526.

\bibitem{KET} K.-E. Thylwe (to be published).

\end{thebibliography}
\end{document}